# Evolution of Entanglement in Groverian Search Algorithm: n-qudit System

Arti Chamoli and C. M. Bhandari
Indian Institute of Information Technology, Allahabad, Deoghat, Jhalwa, Allahabad-211011, India.
Email: achamoli@iiita.ac.in, cmbhandari@yahoo.com

Entanglement plays a crucial role in quantum processes particularly those pertaining to quantum information and computation. An analytic expression for an entanglement measure defined in terms of success rate of Grover's search algorithm has been obtained for qutrit systems with real coefficients and the calculated results agree well with the conventional entropy based measure. The entanglement measure is further generalized for qudit (d-level) systems with real coefficients.

**PACS**
03.65.Ud, 03.67.–a,

## Introduction

Quantum entanglement is the heart and soul of quantum information processing. Its role is distinct and explicit in some situations such as teleportation and superdense coding, whereas it is not so explicit in search algorithms. For obvious reasons most research effort so far has been focused on quantum bits or qubits. The simplicity of two-state systems and the ease with which they can be handled has been primarily responsible for this. However, in principle there is no reason to limit quantum information and computation architectures to two-level systems. The total dimensionality of Hilbert space can be increased by considering qudits; a d-level quantum system which takes d=2 for qubits. The next reasonable step is to look for quantum computation architectures with d=3 which is a qutrit or a three level qudit. During recent years several researchers have investigated such systems in different contexts. Generation and characterization of entanglement for three level system[1]; quantum key distribution protocol with qutrits[2]; quantum tomography for qudits[3]; entanglement swapping between multi qudits[4]; discrimination among Bell states of qudits[5]; GHZ paradox for many qudits[6]; quantum computing with qudits[7]; quantum communication complexity protocol with two entangled qutrits[8]; bounds of entanglement between qudits[9]; entanglement among qudits[10] are worth mentioning in this context. All this unambiguously marks entanglement as the marrow of the theory of information and computation. Thus quantification of the same is of utmost importance. Consequently many entanglement measures have been proposed for qubits to date [11-18]. Recently, F Pan et al [19] presented a classification for entangled bipartite qutrit states based on an entanglement measure [20]. In [21] authors have obtained an entanglement measure for certain kind of four qubit states. Entanglement in a real four qubit system was expressed in terms of its success probability as the initial state of Grover's search algorithm. The measure meets the requirements of being zero for a product state and being invariant under local unitary transformations. Following the same line of thought a measure of entanglement has been framed for qutrit systems. A qutrit is a unit of quantum information whose substates exist in a three dimensional Hilbert space. Accordingly, an n-qutrit system can have $3^n$ subsystems. Violation of local realism being escalated in qutrit correlations enhances its suitability for tasks like cryptography. Use of qutrit systems makes the quantum cryptography protocols robust against eavesdropping attack [2, 22-23]. In addition, security of quantum bit commitment and coin flipping protocols is higher with entangled qutrits[24]. All these breakthroughs have motivated the development of an entanglement measure for d-dimensional systems, where d > 2. A qutrit system is of special interest because information processing appears to have great potential in a three-level system as it best fits into dimensionality aspect of Hilbert space [25]. The Hilbert space dimensionality is maximized for d=3, and hence the computing power.

**Grover's search algorithm and entanglement measure**

We start with a brief review of modified unsorted database search [26], thereby expressing entanglement measure as a derivative of search algorithm. Search space for n qubits has $N$ elements such that $N = 2^n$. Thus the elements can be represented by an n qubit register. Out of these $N$ elements a subset of r elements is marked and we wish to search the whole space in order to find a marked element. The state of n qubits is given by the state $|\phi\rangle$. The search algorithm proceeds with the introduction of an ancilla qubit $|0\rangle_q$ along with the input register $|\phi\rangle$, in the following way:

1. A product of arbitrary local operations, $V = U_1 \otimes U_2 \otimes \ldots \otimes U_n$ on the register and the gate HX on the the ancilla qubit is applied.
$$V|\phi\rangle \otimes HX|0\rangle_q$$
(1)

2. Now the marked state is rotated by a phase of $\pi$ radians. Next all register states are rotated by $\pi$ radians around the average amplitude of the register state.

These two operations constitute Grover iteration $U_G$. The Grover iterations are applied m times, till the amplitude of the marked state reaches a maximum value.

3. Finally, the register is measured in the computational basis. If $P_{max}$ is the maximal success probability of search algorithm where maximization is over all possible local unitary operations in the initial step, then $P_{max}$ can be written in terms of $U_G^m$ (i.e. m Grover iterations). $P_{max}$ is then obtained by averaging uniformly over all $N$ possible values for s ($|s\rangle$ being the marked state).

Thus,
$$P_{max} = \max_{U_1 \cdots U_n} \frac{1}{N} \sum_{s=0}^{N-1} \left|\langle s|U_G^m (U_1 \otimes U_2 \otimes \cdots \otimes U_n)|\phi\rangle\right|^2$$
(2)

For a general state, we consider the effect of the Grover iterations on an uniform superposition state $|\eta\rangle = \sum_x \frac{|x\rangle}{\sqrt{N}}$. Applying m Grover iterations to this state yields:

$$U_G^m|\eta\rangle = |s\rangle + O\left(\frac{1}{\sqrt{N}}\right)$$
(3)

The second term is a small correction term.

Multiplying by $\left(U_G^m\right)^\dagger$ and taking the hermitian conjugate gives

$$\langle s|U_G^m = \langle \eta| + O\left(\frac{1}{\sqrt{N}}\right)$$

(4)

Substituting eq.(4) in eq.(2) gives

$$P_{max} = \max_{U_1----U_n} \frac{1}{N}\sum_{s=0}^{N-1}\left|\langle \eta|U_1 \otimes U_2 \otimes ---- \otimes U_n|\phi\rangle\right|^2 + O\left(\frac{1}{\sqrt{N}}\right)$$

(5)

$|\eta\rangle$ being a product state implies that $U_1^\dagger \otimes U_2^\dagger \otimes ---- \otimes U_n^\dagger|\eta\rangle$ is also a product state. Thus optimization can be considered over product states. Hence,

$$P_{max} = \max_{|e_1----e_n\rangle}\left|\langle e_1 ---- e_n|\phi\rangle\right|^2 + O\left(\frac{1}{\sqrt{N}}\right)$$

(6)

$|e_1\rangle \otimes --- |e_n\rangle$ being a product state of n qubits.

If the input state were a product state, only then would $P_{max}$ be one up to some small corrections. This suggests that $P_{max}$ is affected by the entanglement of the initial state $|\phi\rangle$.

Hence based on the maximum success probability $P_{max}$, an entanglement measure was framed in [26, 27]. This was followed by authors in [21] to study entanglement for a four qubit system. In the present letter, the same approach has been extended to three level systems. An analytical expression giving entanglement measure for qutrit system is derived from modified Grover's search algorithm. The generalized expression for qudit (d >3) systems has also been derived.

According to the definition [26], the entanglement measure $G(\psi)$ in terms of maximum success probability $P_{max}$ can be expressed as

$$G(\psi) = \sqrt{1 - P_{max}}$$

(7)

Dependence on $P_{max}$ makes it obvious that $G(\psi)$ will have its values in the range $0 \leq G(\psi) \leq 1$.

**Groverian Entanglement Measure For Qutrit Systems**
A qutrit being a three-level (d=3) system can be expressed as

$$|\xi\rangle = a_1|1\rangle + a_2|2\rangle + a_3|3\rangle$$

where $a_1, a_2, a_3 \neq 0$ and satisfy the normalization condition $|a_1|^2 + |a_2|^2 + |a_3|^2 = 1$ $|1\rangle, |2\rangle, |3\rangle$ thereby forming an orthonormal basis for a qutrit. Representation in a three dimensional Hilbert space allows an n-qutrit system to have $3^n$ different states simultaneously.

A pure state $|\psi\rangle$ of n-qutrits can be expressed as

$$|\psi\rangle = \sum_{i=1}^{3^n} a_i |i\rangle \qquad (8)$$

where

$$|i\rangle = |i_1, i_2, \ldots i_n\rangle \qquad (9)$$

We aim to obtain an entanglement measure in terms of $P_{\max}(\psi)$. For the same we consider a general product state of n-qutrits:

$$|e\rangle = |e_1\rangle \otimes |e_2\rangle \otimes ----- \otimes |e_n\rangle. \qquad (10)$$

A single qutrit ignoring global phases can be written as

$$|e_k\rangle = e^{i\chi_k} \sin\theta_k \cos\gamma_k |1\rangle_k + e^{i\chi'_k} \sin\theta_k \sin\gamma_k |2\rangle + \cos\theta_k |3\rangle_k \qquad (11)$$

where $k = 1, 2, \ldots, n$, $0 \leq \chi_k, \chi'_k \leq 2\pi$, $0 \leq \theta_k, \gamma_k \leq \pi/2$ and $0 \leq \chi_k, \chi'_k \leq 2\pi$.

The product state $|e\rangle$ can thus be written as
$$|e\rangle = |e_1\rangle \otimes ----- \otimes |e_n\rangle$$
$$= e^{i\chi_1} \sin\theta_1 \cos\gamma_1 --- e^{i\chi_n} \sin\theta_n \cos\gamma_n |1----1\rangle + e^{i\chi_1} \sin\theta_1 \cos\gamma_1 --- e^{i\chi'_n} \sin\theta_n \sin\gamma_n |1---2\rangle + --$$
$$+ ----- + \cos\theta_1 --- \cos\theta_n |3-----3\rangle \qquad (12)$$

Therefore,

$$|e\rangle = \sum_{i=1}^{3^n} c_i |i\rangle \qquad (13)$$

where $|i\rangle = |i_1, i_2, \ldots i_n\rangle$ and

$$c_i = \prod_{k=1}^{n} \left(e^{i\chi_k} \sin\theta_k \cos\gamma_k\right)^{i_k=1} \left(e^{i\chi'_k} \sin\theta_k \cos\gamma_k\right)^{i_k=2} \left(\cos\theta_k\right)^{i_k=3} \qquad (14)$$

The overlap between the initial state $|\psi\rangle$ and product state $|e\rangle$, given by $\langle e|\psi\rangle$, is utilized to obtain Groverian measure of entanglement. This incorporates the maximization of the function

$$P(\theta_1--\theta_n,\gamma_1--\gamma_n,\chi_1--\chi_n,\chi_1'--\chi_n',\psi)=|\langle e|\psi\rangle|^2 \tag{15}$$

with respect to variables $\theta_k, \gamma_k, \chi_k, \chi_k'$, and k = 1-----n.

The maximum success probability thus becomes

$$P_{max}(\psi) = \max_{\substack{\theta_1---\theta_n,\gamma_1---\gamma_n,\\ \chi_1---\chi_n,\chi_1'---\chi_n'}} P(\theta_1---\theta_n,\gamma_1---\gamma_n,\chi_1---\chi_n,\chi_1'---\chi_n',\psi) \tag{16}$$

up to a correction term of order $1/\sqrt{N}$ and the range of maximization is $0 \le \theta_k \gamma_k \le \pi/2$ and $0 \le \chi_k \chi_k' \le 2\pi$. $P_{max}(\psi)$ can be obtained by maximizing P in eq.(15) with respect to variations in $\theta_k, \gamma_k, \chi_k$, and $\chi_k'$ and equating them to zero i.e.,

$$\frac{\partial P}{\partial \theta_k} = \frac{\partial P}{\partial \gamma_k} = \frac{\partial P}{\partial \chi_k} = \frac{\partial P}{\partial \chi_k'} = 0, \tag{17}$$

for k = 1,2,-----n.

For simplicity, the product state $|e\rangle$ with real amplitudes is considered, i.e., all the $\chi_k$ and $\chi_k'$ are 0 or $\pi$, leading to $\exp(i\chi_k) = \exp(i\chi_k') = \pm 1$. This can be removed by doubling the range of $\theta_k$ to $-\pi/2 \le \theta_k \le \pi/2$, thus $\sin\theta_k$ can be both positive and negative for the same value of $\cos\theta_k$.

**A Two-Qutrit System**
The formulation outlined above is applied here to a two qutrit system. A pure two qutrit quantum state $|\psi\rangle$ is written as

$$|\psi\rangle = \sum_{i=1}^{3}\sum_{j=1}^{3} a_{ij}|ij\rangle \tag{18}$$

Hence an overlap of $|\psi\rangle$ with a general product state of two single qutrits, as given by eq.(11) is

$P(\theta_1,\theta_2,\gamma_1,\gamma_2,\psi) = |\langle e|\psi\rangle|^2 =$
$[a_{33}\cos\theta_1\cos\theta_2 + (a_{31}\cos\gamma_2 + a_{32}\sin\gamma_2)\cos\theta_1\sin\theta_2 + (a_{13}\cos\gamma_1 + a_{23}\sin\gamma_1)\sin\theta_1\cos\theta_2 +$

$$\left(a_{11}\cos\gamma_1\cos\gamma_2 + a_{12}\cos\gamma_1\sin\gamma_2 + a_{21}\sin\gamma_1\cos\gamma_2 + a_{22}\sin\gamma_1\sin\gamma_2\right)\sin\theta_1\sin\theta_2\Big]^2$$

(19)

This can be simplified by assuming $\gamma_1+\gamma_2=\gamma_x$, $\gamma_1-\gamma_2=\gamma_y$, $\theta_1+\theta_2=\theta_x$ and $\theta_1-\theta_2=\theta_y$,

$$P(\theta_x,\theta_y,\gamma_1,\gamma_2,\gamma_x,\gamma_y)=$$

$$\frac{1}{4}\Bigg[\bigg\{a_{33}-\bigg(\frac{a_{11}-a_{22}}{2}\cos\gamma_x+\frac{a_{11}+a_{22}}{2}\cos\gamma_y+\frac{a_{21}+a_{12}}{2}\sin\gamma_x+\frac{a_{21}-a_{12}}{2}\sin\gamma_y\bigg)\bigg\}\cos\theta_x$$

$$+\bigg\{a_{33}+\bigg(\frac{a_{11}-a_{22}}{2}\cos\gamma_x+\frac{a_{11}+a_{22}}{2}\cos\gamma_y+\frac{a_{21}+a_{12}}{2}\sin\gamma_x+\frac{a_{21}-a_{12}}{2}\sin\gamma_y\bigg)\bigg\}\cos\theta_y$$

$$+\{a_{13}\cos\gamma_1+a_{23}\sin\gamma_1+a_{31}\cos\gamma_2+a_{32}\sin\gamma_2\}\sin\theta_x$$

$$+\{a_{13}\cos\gamma_1+a_{23}\sin\gamma_1-a_{31}\cos\gamma_2-a_{32}\sin\gamma_2\}\sin\theta_y\Bigg]^2$$

(20)

Then by solving $\dfrac{\partial P}{\partial \theta_x}=\dfrac{\partial P}{\partial \theta_y}=\dfrac{\partial P}{\partial \gamma_x}=\dfrac{\partial P}{\partial \gamma_y}=\dfrac{\partial P}{\partial \gamma_1}=\dfrac{\partial P}{\partial \gamma_2}=0$, maximum is found for $\theta_x,\theta_y,\gamma_x,\gamma_y,\gamma_1$ and $\gamma_2$. The values obtained are substituted in eq.(20).

$P_{\max}(\psi)$ is finally as:

$$P_{\max}(\psi)=\frac{1}{4}\Bigg[\bigg[\bigg\{a_{33}-\frac{1}{2}\bigg(\sqrt{(a_{11}-a_{22})^2+(a_{21}+a_{12})^2}+\sqrt{(a_{11}+a_{22})^2+(a_{21}-a_{12})^2}\bigg)\bigg\}^2+$$

$$\bigg\{\sqrt{a_{13}^2+a_{23}^2}+\sqrt{a_{31}^2+a_{32}^2}\bigg\}^2\bigg]^{\frac{1}{2}}+\bigg[\bigg\{a_{33}+\frac{1}{2}\bigg(\sqrt{(a_{11}-a_{22})^2+(a_{21}+a_{12})^2}+\sqrt{(a_{11}+a_{22})^2+(a_{21}-a_{12})^2}\bigg)\bigg\}^2$$

$$+\bigg\{\sqrt{a_{13}^2+a_{23}^2}-\sqrt{a_{31}^2+a_{32}^2}\bigg\}^2\bigg]^{\frac{1}{2}}\Bigg]^2$$

(21)

Eventually, Groverian entanglement of a state $|\psi\rangle$ can be calculated by

$$G(\psi)=\sqrt{1-P_{\max}(\psi)}$$

This measure can quantify the entanglement present in various two qutrit states. Also it ca very well categorize an entangled and a product state. As for product states

$P_{\max}(\psi) = 1$, whereas it is never so for any entangled state. The importance of an entanglement measure lies in the fact that variation in the amount of entanglement in quantum states affects quantum computation and information processing.

For a product state, it is easy to find out $P_{\max}(\psi) = 1$, which leads to zero entanglement. This is in correspondence with one of the criteria of an entanglement measure which states $G(\psi)$ should vanish for product states.

$P_{\max}(\psi)$ and finally Groverian entanglement is calculated below for certain two-qutrit systems.

For a maximally entangled state of the type
$$|\psi\rangle = \frac{1}{\sqrt{3}}(|11\rangle + |22\rangle + |33\rangle)$$
$P_{\max}(\psi) = 0.3333$, and $G(\psi) = 0.8165$.

For another extremally entangled state,
$$|\psi\rangle = \frac{1}{\sqrt{2}}(|11\rangle + |22\rangle)$$
$P_{\max}(\psi) = 0.5$ and thus $G(\psi) = 0.7071$. For the same state the value of entanglement as reported in [19] is 0.63093.

We now turn to the general qudits with an arbitrary $d \geq 3$. Consider a pure state
$$|\Phi\rangle = \sum_{i=1}^{N^n} a_i |i\rangle \quad (22)$$

of $n$-qudits, where,
$$|i\rangle = |i_1, i_2, \ldots i_n\rangle \quad (23)$$

The maximum success probability of Grover's search algorithm is given by eq.(6). To obtain an explicit expression for the same, a general product state of $n$-qudits can be written as:
$$|E\rangle = |E_1\rangle \otimes \ldots \otimes |E_n\rangle \quad (24)$$

The single qudit state can be represented by

$$|E_j\rangle = e^{i\chi_{1j}} \sin\alpha_{1_j} \prod_{k_j=2}^{N-1} \cos\alpha_{k_j} |1\rangle + \sin\alpha_{1_j} \sum_{\substack{x=2\\m=2}}^{N-1} e^{i\chi_{x_j}} \sin\alpha_{m_j} \prod_{k=2}^{m-1} \cos\alpha_{k_j} |K\rangle + \cos\alpha_{1_j} |N\rangle$$

where $|K\rangle = |2\rangle .... |N-1\rangle$ terms.

(25)

where $j = 1, 2, ......., n$. Ignoring global phases, the range of $\alpha_{k_j}$ is $0 \le \alpha_{k_j} \le \pi/2$ and that of $\chi_{x_j}$ is $0 \le \chi_{x_j} \le 2\pi$.

The product state of $n$-qudits thus can be written as

$$|E_1\rangle \otimes ....... \otimes |E_n\rangle = e^{i\chi_{1_1}} \sin\alpha_{1_1} \prod_{k_1=2}^{N-1} \cos\alpha_{k_1} ....... e^{i\chi_{1_n}} \sin\alpha_{1_n} \prod_{k_n=2}^{N-1} \cos\alpha_{k_n} |1.......1\rangle$$

$$+ e^{i\chi_{1_1}} \sin\alpha_{1_1} \prod_{k_1=2}^{N-1} \cos\alpha_{k_1} ...... \sin\alpha_{1_n} e^{i\chi_{2_n}} \sin\alpha_{2_n} |1.......2\rangle$$

$$+ ....... + \cos\alpha_{1_1} ..... \cos\alpha_{1_n} |N......N\rangle$$

(26)

The product state can be written as

$$|E\rangle = \sum_{i=1}^{N^n} C_i |i\rangle$$

(27)

where the coefficient of the basis state $|i\rangle = |i_1, i_2, ....i_n\rangle$ is

$$C_i = \prod_{j=1}^{n} \left( e^{i\chi_{1j}} \sin\alpha_{1_j} \prod_{k_j=2}^{N-1} \cos\alpha_{k_j} \right)^{i_k=1} \left( \sin\alpha_{1_j} \sum_{\substack{x=2\\m=2}}^{N-1} e^{i\chi_{x_j}} \sin\alpha_{m_j} \prod_{k=2}^{m-1} \cos\alpha_{k_j} \right)^{i_k=2....N-1} \left( \cos\alpha_{1_j} \right)^{i_k=N}$$

(28)

Thus the Groverian measure of entanglement for $n$-qudits can be obtained by maximizing the function

$$P(\alpha_{1_1}, ......, \alpha_{k_n}, \chi_{1_1}, ....., \chi_{x_n}, \Phi) = |\langle E_1, ....E_n | \Phi \rangle|^2$$

(29)

with respect to $\alpha_{k_j}, \chi_{x_j}, j = 1, ....., n$. The maximum probability can now be written as

$$P_{\max}(\Phi) = \max_{\alpha_{1_1}, ......, \alpha_{k_n}, \chi_{1_1}, ....., \chi_{x_n}} P(\alpha_{1_1}, ......, \alpha_{k_n}, \chi_{1_1}, ....., \chi_{x_n}, \Phi)$$

(30)

up to a correction term of order $1/\sqrt{N^n}$, and the maximization is taken in the range $0 \leq \alpha_{k_j} \leq \pi/2$ and $0 \leq \chi_{x_j} \leq 2\pi$.

**Conclusion**
We have generalized an operational entanglement measure based on the success rate of Grover's search algorithm in reaching the desired rate to $n$-qudits system. A quantum system belonging to a 3-D Hilbert space is an optimal one [25] in view of quantum computation and communication. Thus quantification of entanglement for such system facilitates its use in various information processing tasks. The expression for $P_{\max}(\psi)$ as obtained above easily gives the value of $G(\psi)$ just by inserting the value of normalized coefficients in the expression. The value of $G(\psi)$ as calculated for certain states are on expected lines. The expression for $P_{\max}(\psi)$, being conceptually based on the success probability of Grover's unsorted database search, indicates that entanglement is generated, rises to a maximum, and then finally vanishes during the processing of Grover's algorithm. Same can be exemplified by a two qutrit product state as the initial state of the search algorithm. Unlike a two qubit product state as the starting state for the search algorithm to proceed, a two qutrit product state has maximum probability of reaching the desired state after two iterations. Entanglement is zero for the initial state, then rises to a certain value after first iteration and finally decays after second iteration. The starting state $|\psi\rangle$, a product of two qutrits in uniform superposition is

$$|\psi\rangle = \frac{1}{3}(|11\rangle + |12\rangle + |13\rangle + |21\rangle + |22\rangle + |23\rangle + |31\rangle + |32\rangle + |33\rangle)$$

Operators $P_W = 1 - 2|W\rangle\langle W|$, $|W\rangle$ being the desired state and then $P_\psi = 2|\psi\rangle\langle\psi| - 1$ (the combination of the two constitutes one Grover iteration) are applied to $|\psi\rangle$. Two Grover iterations complete the search process. The analytical expression for $P_{\max}(\psi)$ also verifies that if the initial state of Grover's search algorithm is an entangled one then the performance of search algorithm is deteriorated. Entanglement being a vital part of quantum computation and communication can be exploited to the fullest only if its value is known for any quantum state under consideration. The expression for $P_{\max}(\psi)$ derived above solves the purpose well for a two qutrit system.

Figure1. shows the evolution of entanglement as the Grover's search algorithm proceeds. The solid line shows the quantification of entanglement as calculated from the expression for $P_{\max}(\psi)$ whereas the dotted line quantifies entanglement on the basis of entropy of entanglement. This quantification is done by the Von Neumann entropy with

$$S_\psi = -Tr\left[(\rho_\psi)_i \log_3 (\rho_\psi)_i\right]$$

where i = 1 or 2 for a two qutrit system, and $(\rho_\psi)_i$ (i = 1 or 2) is the reduced density matrix with particle 2 or 1, respectively, traced out. Degree of entanglment has been calculated for all the intermediate states of two qutrit search algorithm generated after applying $P_W$ and $P_\psi$ operators to the initial state $|\psi\rangle$ within two grover iterations.

**ACKNOWLEDGEMENT**
Authors are thankful to Dr. M. D. Tiwari for his keen interest and support. Arti Chamoli is thankful is thankful to IIIT, Allahabad for financial support.

**REFRENCES:**
[1]  C. M. Caves and G. J. Milburn, Opt. Comm. **179**, 439 (2000).
[2]  D. Bruss and C. Macchiavello, Phys. Rev. Lett. **88**, 127901 (2002).
[3]  R. T. Thew, K. Nemoto, A. G. White, and W. J. Munro, quant-ph/0201052.
[4]  J. Bouda and V. Buzek, Journal of Phys. A Math. Gen. **34**, 4301 (2001).
[5]  M. Dusek, quant-ph/0107119.
[6]  N. J. Cerf, S. Massar, and S. Pironio, quant-ph/0107031.
[7]  S. D. Bartlett, H. de Guise, and B. C. Sanders, quant-ph/0109066.
[8]  C. Brukner, M. Zukowski, and A. Zeilinger, quant-ph/0205080.
[9]  V. M. Kendon, K. Zyczkowski, and W. J. Munro, quant-ph/0203037.
[10] P. Rungta, W. J. Munro, K. Nemoto, P. Deuar, G. J. Milburn, and C. M. Caves, quant-ph/0001075.
[11] V. Vedral, M. B. Plenio, M. A. Rippin, and P. L. Knight, Phys. Rev. Lett. **78**, 2275 (1997).
[12] S. Hill and W. K. Wootters, Phys. Rev. Lett. **78**, 5022 (1998).
[13] W. K. Wootters, Phys. Rev. Lett. **80**, 2245 (1998).
[14] V. Vedral and M. B. Plenio, Phys. Rev. A **57**, 1619 (1998).
[15] V. Coffman, J. Kundu, and W. K. Wootters, Phys. Rev. A **61**, 052306 (2000).
[16] A. Wong and N. Christensen, Phys. Rev. A **63**, 044301 (2001).
[17] G. Vidal and R. F. Werner, Phys. Rev. A **65**, 032314 (2002).
[18] W. D¨ur, G. Vidal, and J. I. Cirac, Phys. Rev. A **62**, 062314 (2000).
[19] F. Pan, Guoying Lu, and J. P. Drayer, quant-ph/0510178.
[20] F. Pan, D Liu, G. Y. Lu, and J. P. Drayer, Int. J. Theor. Phys. **43**, 1241 (2004).
[21] A. Chamoli and C. M. Bhandari, Phys. Lett. A **346,** 17 (2005).
[22] N. J. Cerf, M. Bourennane, A. Karlsson, and N. Gissin, Phys. Rev. Lett. **88**, 127902 (2002).
[23] T. Durt, N. J. Cerf, N. Gissin, and M. Zukowski, Phys. Rev. A **67**, 012311 (2003).
[24] R. W. Spekkens and T. Rudolph Phys. Rev. A **65**, 012310 (2002).
[25] A. D. Greentree, S. G. Schirmer, F Green, L. C. L. Hollenberg, A. R. Hamilton and R. G. Clark, Phys. Rev. Lett. **92**, 097901 (2004).
[26] O. Biham, M. A. Nielsen, and T. Osborne, Phys. Rev. A. **65**, 062312 (2002).
[27] Y. Shimoni, D. Shapira, and O. Biham, Phys. Rev. A. **69**, 062303 (2004).
[28] R. Das, A. Mitra, V. Kumar S, and A. Kumar, quant-ph/0307240.
[29] A. B. Klimov, R. Guzmn, J. C. Retamal, and C. Saavedra, Phys. Rev. A **67**, 062313 (2003).

[30] D. Mc Hugh and J. Twamley, New J. Phys. **7**, 174 (2005).
[31] R. T. Thew, A. Acin, H. Zbinden, and N. Gisin, Quantum Information and Computation, **Vol.4**, No.2, 93 (2004).
[32] Y. I. Bogdanov, M. Chekhova, S. Kulik, G. Maslennikov, C. H. Oh, M. K. Tey, and A. Zhukhov, Phys. Rev. Lett., **93**, 230503 (2004).

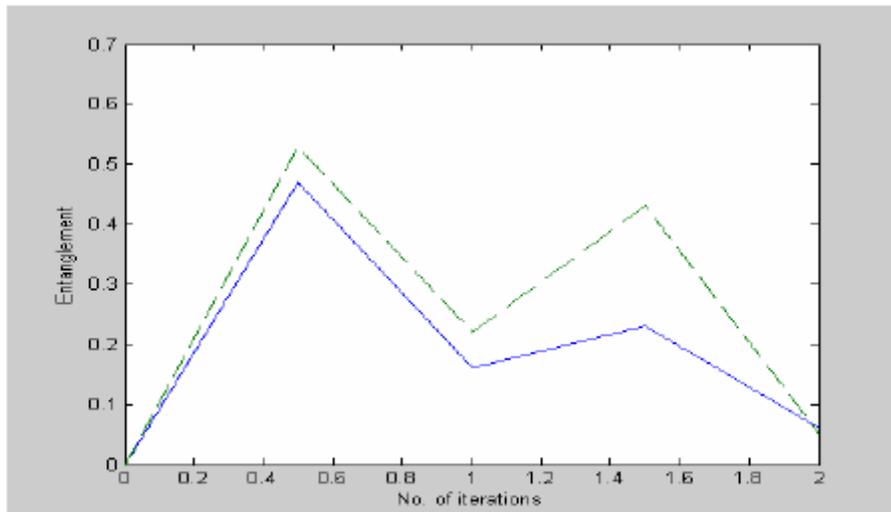

FIG1. Evolution of entanglement for a two qutrit system with application of Grover operators. Solid line indicates Groverian measure; dotted line indicates entanglement measure as entropy of entanglement.